\documentclass[twocolumn,prl,aps,letterpaper]{revtex4}
\usepackage{graphics}
\usepackage{epsfig}
\usepackage{natbib}

\begin{document}
\newcommand*{\ket}[1]{$|{#1}\rangle$}
\newcommand*{\beq}{\begin{equation}}
\newcommand*{\eeq}{\end{equation}}

\title{ Matter-Wave Decoherence due to a Gas Environment in an Atom
Interferometer}

\author{Hermann Uys}
\author{John D. Perreault}
\author{Alexander D. Cronin}
\affiliation{Department of Physics, University of Arizona, Tucson,
AZ 85721}

\begin{abstract}

Decoherence due to scattering from background gas particles is
observed for the first time in a Mach-Zehnder atom interferometer,
and compared with decoherence due to scattering photons.  A single
theory is shown to describe decoherence due to scattering either
atoms or photons.  Predictions from this theory are tested by
experiments with different species of background gas, and also by
experiments with different collimation restrictions on an atom
beam interferometer.
\end{abstract}
\date{\today}
\maketitle

When a quantum system in a superposition of states interacts with
an environment, the coherence of the superposition can be lost.
Modern decoherence theory explains this is a result of
entanglement between the system and unobserved degrees of freedom
in the environment \cite{ZUR91}. Understanding how different
environments cause decoherence is important for applications such
as atom interferometry or quantum computation, where coherent
superpositions are required.

In this Letter we compare two different mechanisms for
decoherence: scattering atoms and scattering light. A dilute gas
of massive particles and a beam of radiation are quite different
environments, yet they both cause contrast loss in our atom
interferometer. The data and analysis presented here shows that
\emph{gas decoherence} (atom scattering) and \emph{photon
decoherence} (light scattering) can be understood with a single
universal theory described by Tan and Waals \cite{TAW93} and also
Tegmark \cite{TEG93}. Motivated by this theory, we show that the
distribution of momentum transfer from the environment to the
\emph{detected} atoms determines the amount of contrast loss
regardless of the kind of objects being scattered in the
environment.

To study gas decoherence we vary the background pressure in the
entire interferometer chamber whilst monitoring the transmitted
flux, as well as the interference contrast which we define as $C =
(I_{max}-I_{min})/(I_{max}+I_{min})$, where  $I_{max(min)}$ is the
maximum (minimum) count rate in the interference fringes. The
presence of contrast reflects the fact that the atoms are in a
coherent superposition of states. A related issue is the
attenuation of atom beam intensity by the environments. The
attenuation is the same on both paths of the interferometer and is
therefore the equal to the attenuation of the average detected
intensity, $\langle I \rangle = (I_{max} + I_{min})/2$. In what
follows we will denote the contrast and intensity in the absence
of scattering by $C_0$ and $I_0$. Because the room temperature
background gas atoms have more momentum than photons by a factor
of $\simeq 10^5$, they can deflect beam atoms by large enough
angles that they often miss our detector.  As a result, the gas
environment reduces the average atom beam intensity to 10\% before
the contrast is halved. By comparison photon scattering does not
reduce the detected flux (see Fig. 1).

\begin{figure}
\begin{center}
\includegraphics[scale=0.55]{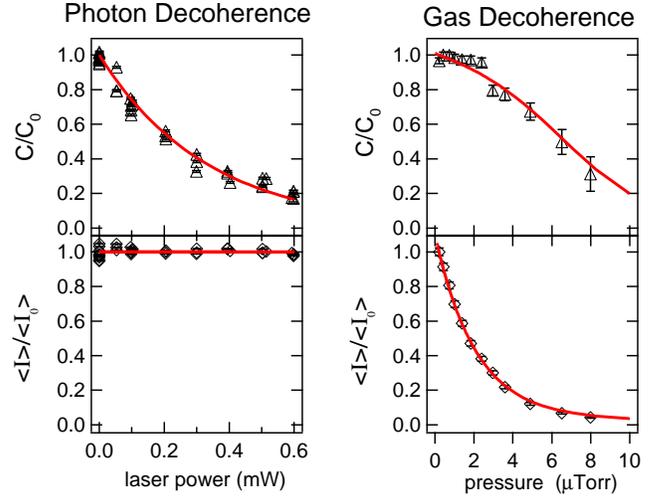}
\end{center}
\caption{Comparison of photon decoherence (left) to gas
decoherence (right).  Contrast and Intensity in an atom beam
interferometer are reported as a function of resonant laser beam
power or background gas pressure.  The curves come from a general
theory of decoherence given by Equations (2) and
(3).}\label{compare}
\end{figure}

One can compare the two environments as two gedanken microscopes
that use either light waves or de Broglie waves of background gas
to detect which path atoms took in the interferometer. According
to Feynmann's Heisenberg microscope idea \cite{FLS65}, a quantum
system can be localized with a spatial resolution $\delta x$ of
\begin{equation}\delta x \geq \frac{\lambda}{2 \mbox{ NA}} =
\frac{h}{2p\mbox{ NA}},\label{eq:resolution}\end{equation} where
$\lambda$ is the wavelength of the light or the wavelength of the
gas particles used to make the microscope, $p$ is the momentum of
the photons (or gas particles), and $\mbox{NA}$ is the numerical
aperture of the microscope. The inequality in Eq.
(\ref{eq:resolution}) becomes an equals sign if the microscope is
limited only by diffraction. Heuristically, by detecting an atom's
position the microscope can observe the particle nature of an
atom, which is complementary to the wave nature.  Hence, if such a
microscope can even in principle resolve which path each atom took
in the interferometer, then evidence for the wave nature of atoms
(i.e. the interference fringe contrast) should be unobservable.

Since the gedanken gas microscope has a smaller wavelength probe
than the light microscope, it is natural to expect that gas
scattering should cause more contrast loss than photon scattering.
In apparent contradiction to this simple picture, the contrast is
not destroyed even when there is a sufficient gas pressure to
attenuate the atom beam. As we shall explain, the analysis of the
gedanken microscope resolution must be adjusted to include the
probability distribution of momentum kicks from the environment to
each detected atom, and in the case of the gas environment this
lead to poorer gedanken microscope resolution than predicted by
the ideal limit.
\begin{figure}
\centering
\begin{center}
\includegraphics[scale=0.4]{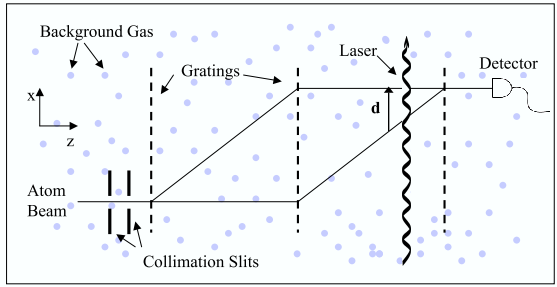}
\end{center}
\caption{Schematic of the atom interferometer and scattering
scenarios. For gas decoherence the entire interferometer is
exposed to background gas, allowing scattering to take place
anywhere along both arms of the interferometer. Previous
experiments \cite{CHL95,BER97,KCR01} studied photon scattering
from the two arms at a location where the separation vector
$\mathbf{d}$ is defined.  \label{interf}}
\end{figure}

Our experiment employs a Mach-Zehnder atom interferometer, shown
in Fig. \ref{interf}, with two arms formed by the zeroth and first
diffraction orders of a supersonic Na atom beam (mean velocity
3000 m/s and $\Delta v/\langle v \rangle\approx 1/10$) that passes
through a 100-nm period grating.  A second grating redirects the
beams so they overlap and make flux density interference fringes
at the position of a third grating. Atoms transmitted through the
third grating are detected with a hot wire, and oscillations in
the flux due to the interference contrast are observed when the
third grating is translated. At a background gas pressure of
$2\times10^{-7}$ Torr, the contrast is $C_0\simeq 25\%$ and the
average detected atom flux is $I_0\simeq 100,000$ counts per
second. The beam is collimated to 2$\times 10^{-5}$ radians by two
10 $\mu$m slits separated by 1 m, and the detector is $50$ $\mu$m
in diameter. The gratings are each separated by 1 m.  To study gas
decoherence we vary the background pressure by controlling a gas
leak into the interferometer chamber whilst monitoring the beam
intensity and interference contrast. Note that the entire
interferometer chamber is filled with gas so scattering can take
place anywhere along either arm of the interferometer.  This setup
is different from atom interferometer experiments in which a gas
cell was placed on one arm \cite{SCE95}, or neutron interferometer
experiments with an absorbing structure in one arm
\cite{neutrons}.  Those cases lead to attenuation in one arm only.
In our case both arms are exposed to the same gas environment and
experience the same attenuation.  (We will discuss gas decoherence
in relation to \cite{SCE95,neutrons} more in the conclusion).


Our apparatus was previously used to study photon decoherence
\cite{CHL95,BER97,KCR01,PCK01}. That experiment used a laser beam
tuned to the $\lambda$ = 590 nm transition of the Na beam atoms
that was positioned as indicated in Fig.~\ref{interf}, so as to
scatter off of both arms of the interferometer.  The contrast was
monitored as a function of laser intensity and as a function of
the separation vector, $\mathbf{d}$, between the two arms of the
interferometer at the location of scattering. This Letter is the
first report of gas decoherence in a Mach-Zehnder interferometer.
A Talbot Lau interferometer was used recently to study gas
decoherence \cite{HUB03}, but our observations are in a different
regime since the momentum spread of our detected beam is 10,000
times smaller than in the Talbot Lau interferometer \cite{BJU02}.
Unlike \cite{HUB03} we do not get an exponential decay of contrast
with gas pressure (see Fig. 1). It was also suggested in
\cite{HUB03} that gas decoherence could not be observed in our
Mach-Zehnder interferometer because all the scattered atoms would
miss the detector. Indeed, in the limit of an infinitely narrow
beam and detector, only atoms with zero recoil could be detected.
In that case the momentum transfer, $(p\mbox{ NA}) = 0$, so by Eq.
(\ref{eq:resolution}) a Heisenberg microscope would have poor
resolution and fringe contrast could therefore be preserved. We
attribute the possibility of gas decoherence in our apparatus to
the non-zero size of the atom beam and the detector. To confirm
this we present data showing that gas decoherence depends on the
beam collimation in Fig.~3.

\begin{figure}[t]
\begin{center}
\includegraphics[scale=0.7]{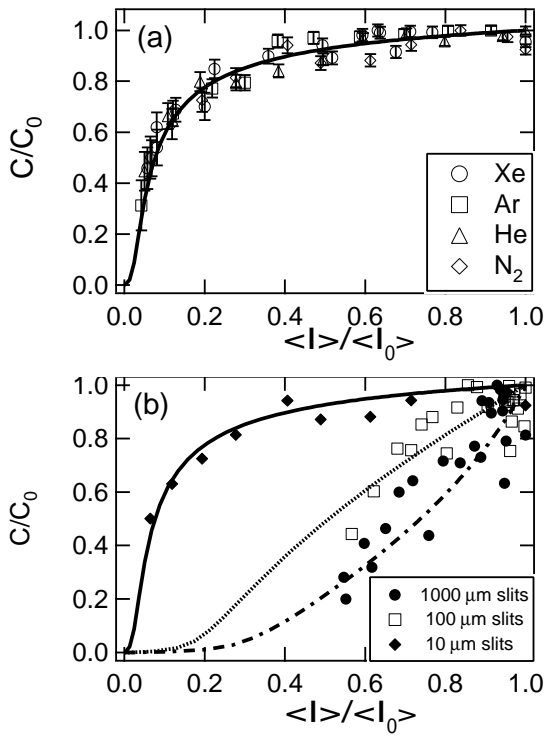} 
\end{center}
\caption{Contrast and intensity shown in parametric plots as a
function of the pressure of the background gas. (a) Various
background gas species provide similar results. (b) Large atom
beams cause more rapid loss of contrast. Theory curves are from
Equations (4) and (5).\label{gasdata}}
\end{figure}

Figure \ref{gasdata}(a) compares the contrast loss, $C/C_0$,
observed as a function of the average beam intensity, $\langle I
\rangle / \langle I_0 \rangle$, when different species of
background gas (Xe, Ar, He, and N$_2$) are introduced to the
interferometer chamber. These data form a universal curve even
though quite different amounts of pressure are needed for each gas
to cause a 50\% reduction in atom beam intensity. Fig.
\ref{gasdata}(b) shows similar data obtained with wider
collimating slits.  The solid and dashed lines come from our
decoherence theory described next.

The results of Fig. 3 can be understood within a widely accepted
picture of decoherence that views every system as a subset of a
larger environment that is also governed by quantum mechanics, but
is not monitored by the observer \cite{ZUR91}.  The result of
coupling to this environment is that the off-diagonal elements of
the density matrix describing the system are damped. This picture
was successfully used to explain the photon decoherence experiment
\cite{CHL95,BER97,KCR01,PCK01} in which case the damping of
off-diagonal elements of the density matrix causes a reduction in
contrast to $C = \beta C_0$.  The \emph{decoherence function},
$\beta$, is
\begin{equation}
\label{beta} \beta(\mathbf{d})=  \frac{ \int^{\Delta
k_{max}}_{\Delta k_{min}}
P(\Delta\mathbf{k})e^{-i(\Delta\mathbf{k})\cdot\mathbf{d}}\;d\Delta\mathbf{k}}{\int^{\Delta
k_{max}}_{\Delta k_{min}} P(\Delta\mathbf{k})\;d\Delta\mathbf{k}},
\end{equation}
as predicted by \cite{TAW93} and \cite{TEG93}. In
Eq.~(\ref{beta}), $P(\Delta\mathbf{k})$ is the probability
distribution to undergo a momentum change $\hbar\Delta\mathbf{ k}$
due to scattering. The limits of integration are determined by
$\Delta\mathbf{k}_{max(min)}$ that still permit the atom beam to
be detected.  The denominator simply expresses the total
probability, $P_{det}$, for a beam atom to be detected. We
emphasize that $\Delta\mathbf{k}_{max(min)}$ depends not only on
the size of the detector, but also on the initial position and
momentum of the beam atoms. To find $\beta$ one must therefore
average over the beam width and height.

To describe $P(\Delta\mathbf{k})$ one must take into account the
probability, $P_n$, that an atom undergoes exactly $n$ scattering
events on its trajectory to the detector plane. $P_n$ is
determined by the background gas pressure or radiation intensity.
$P(\Delta\mathbf{k})$ also depends on the probability that the
atom gets a total momentum kick $\Delta\mathbf{k}$ as a result of
these $n$ scattering events. This is given by the convolution
(which we indicate with $*$) of the absolute value squared of the
scattering amplitude $f \equiv f(\mathbf{\Delta k})$  with itself
$n$ times, because the probability distribution function of two
random variables is the convolution of the constituent
distribution functions. We write \beq P(\Delta \mathbf{k}) = P_0
\delta(\Delta\mathbf{k}) + \left ( P_1 \frac{|f|^2}{\sigma_t} +
P_2 \frac{|f|^2*|f|^2 }{\sigma_t^2} + ...\right ) ,\label{Pn}\eeq
where $\delta ()$ is the Dirac delta function, and $f$ is
explicitly normalized by the total scattering cross-section
$\sigma_t$.

In practice we account for the averaging over beam profile in Eq.
(\ref{beta}) in an approximate way by multipying the term in
brackets on the right in Eq. (\ref{Pn}) (i.e. terms accounting for
multiple scattering) by $(1 + \frac{A_{beam}}{A_{det}})$, where
$A_{beam}$ is the cross-sectional area of the beam and $A_{det}$
the cross-sectional area of the detector. The weight factor
$\frac{A_{beam}}{A_{det}}$ then allows for the possibility that
beam atoms that would have missed the detector in the absence of
scattering, might now be scattered onto the detector.

Up to this point, the theory used in Eqs.~(\ref{beta}) and
(\ref{Pn}) is still general enough to describe both photon
decoherence and gas decoherence. Indeed, this universal model of
decoherence from scattering serves as the basis for all the
theoretical curves presented in Figs. (\ref{compare}) and
(\ref{gasdata}). For the photon environment, $P_n$ and $|f(\Delta
k)|^2$ are determined by atom-photon interactions, e.g., dipole
radiation scattering as discussed in
\cite{CHL95,BER97,KCR01,PCK01}. For the gas environment it can be
shown that the probability to scatter $n$ times after travelling a
distance $z$ through the background gas obeys the Poisson
distribution $P_n =
\frac{z^n}{n!\lambda^n}e^{-\frac{z}{\lambda}}$, where $\lambda$ is
the mean free path. Furthermore $f(\Delta k)$ is the complex
amplitude of an outgoing spherical wave in the Lippman-Schwinger
equation, which in the Born approximation is simply the Fourier
transform of the inter-atomic potential \cite{SAK94}.

A thorough calculation of the decoherence function for the gas
environment should also include an average over initial momentum
states of the background gas.  The main effect of the averaging
procedure described by Russek \cite{RUS60} is to scale the
scattering angle in the lab, $\theta$, with respect to that in the
center of mass frame, $\Theta$, according to $\theta \approx
\frac{m_g}{m_g + m_b}\Theta$ where $m_g$ is the mass of the
background gas atom and $m_b$ the mass of the beam atom. This
approximation is valid when the atom beam speed is large compared
the average speed of the background gas atoms, and it was used in
this analysis to find $f(\mathbf{\Delta k})$.

Since scattering from the background gas can take place anywhere
along the interferometer the decoherence function should also be
averaged over $\mathbf{d}$. However, instead of explicitly
averaging over $\mathbf{d}$ we note that terms in the numerator of
the decoherence function, Eq. (\ref{beta}), are small if they
oscillate rapidly over the range of integration, i.e. when
$\mathbf{d}\cdot\Delta \mathbf{k}> 2$. For gas scattering this
rapid oscillation of the integrand occurs for values of
$\mathbf{d}$ corresponding to scattering a distance of $z \geq 1$
mm from the third (or first) grating. Therefore only scattering
that takes place close to the third (or first) grating gives
significant contributions to $\beta$.  The consequence of
averaging over $\mathbf{d}$, which we indicate with angle
brackets, is therefore that to a good approximation \textit{only
atoms that do not scatter contribute coherently to the
interference pattern}. With this consideration the predicted
contrast based on Eqns. (\ref{beta}) and (\ref{Pn}) reduces to


\begin{equation}
\langle C\rangle_d \approx C_0\left\langle\frac{ P_0 }{
P_{det}}\right\rangle_d, \label{final}
\end{equation} and the detected flux is
\begin{equation}
\langle I \rangle_d \approx I_0 \langle P_{det}\rangle_d\mbox{  }.
\label{final}
\end{equation}
These predictions show good agreement with experimental data as
seen in Figs. \ref{compare} and \ref{gasdata}. Note in particular
that data taken with the Na atom beam and all background gasses
tested (He, Ar, Xe, N$_2$), collapse on the same curve in this
plot. This is due to the fact that the spatial extent of the
interatomic potential is very similar for Na and all the gasses
used (the minimum in $V(r)$ is located near 5 Angstroms for each
of these gasses \cite{TEL79}). As a consequence, the widths of the
scattering probability distributions, $|f(\Delta \mathbf{k})|^2$,
are similar for these gasses.



Eqns. (\ref{beta}), (\ref{Pn}) and the Born approximation for
$f(\Delta k)$, taken together allow us to predict that faster
beams or scattering centers with larger spatial extent to their
potential will cause more contrast loss per attenuation, since
both of these scenarios lead to narrower $P(\Delta\mathbf{k})$ and
hence a larger $P_{det}$. Furthermore, a larger detector or larger
atom beam (hence larger $\Delta k_{max}$) will also cause more
contrast loss, since this would increase the incoherent
contribution to the denominator in Eq. (\ref{beta}) due to
scattering.

We have tested one of these predictions by using a beam with
larger cross-sectional area. A wide beam is beneficial for
applications such as gyroscopes \cite{GLK00} where high flux is
desired, but separated beams are not needed.  Fig.~3(b) compares
the contrast, $C/C_0$,  as a function of atom flux, $\langle I
\rangle / \langle I_0 \rangle$, for for three beam sizes. Note
that contrast is diminished more rapidly for the wider beam as
predicted by Eqs. (4) and (5). Optimizing interferometer
performance for practical applications therefore relies on a
tradeoff between desired intensity and contrast at the achievable
vacuum levels.

In the vocabulary of related work \cite{neutrons}, the gas
environment is a stochastic absorber in the quantum limit.
Furthermore, as predicted by \cite{neutrons} for an environment
such as the gas that interacts with both arms of the
interferometer, if only non-scattered atoms were detected then
$P_{det} = P_0$ and $C/C_0$ would remain 1 even as $\langle I
\rangle / \langle I_0 \rangle$ is reduced to $P_0$.

As a final point we discuss the impact of gas decoherence on
experiments in which the atom wave index of refraction due to a
dilute gas was measured \cite{SCE95}.  
Additional incoherent flux that
hits the detector does not affect the fringe phase, nor does it
change the product $\langle I\rangle C$. Thus the gas decoherence
described here should not influence results reported in
\cite{SCE95}.

In conclusion, we have observed loss of interference contrast in a
Mach-Zehnder atom interferometer as a result of increased
background gas pressure.  Momentum transfer from the gaseous
environment causes decoherence of scattered atoms so that only
atoms that undergo no scattering contribute coherently to the
interference pattern.  These results are explained by a general
theory of decoherence that treats gas scattering and photon
scattering equally. This theory allows us to predict that higher
velocity beams, atoms with longer range potentials, a wider
detector or a wider beam will increase the contrast loss from gas
decoherence. This provides quantitative predictions for
interferometer performance in imperfect vacuum.

This research was supported by an Award from Research Corporation
and the National Science Foundation Grant No PHY-0354947.


\begin{thebibliography}{19}
\expandafter\ifx\csname
natexlab\endcsname\relax\def\natexlab#1{#1}\fi
\expandafter\ifx\csname bibnamefont\endcsname\relax
  \def\bibnamefont#1{#1}\fi
\expandafter\ifx\csname bibfnamefont\endcsname\relax
  \def\bibfnamefont#1{#1}\fi
\expandafter\ifx\csname citenamefont\endcsname\relax
  \def\citenamefont#1{#1}\fi
\expandafter\ifx\csname url\endcsname\relax
  \def\url#1{\texttt{#1}}\fi
\expandafter\ifx\csname
urlprefix\endcsname\relax\def\urlprefix{URL }\fi
\providecommand{\bibinfo}[2]{#2}
\providecommand{\eprint}[2][]{\url{#2}}

\bibitem[{\citenamefont{Zurek}(1991)}]{ZUR91}
\bibinfo{author}{\bibfnamefont{W.~H.} \bibnamefont{Zurek}},
  \bibinfo{journal}{Physics Today} \textbf{\bibinfo{volume}{44}},
  \bibinfo{pages}{36} (\bibinfo{year}{1991});
\bibinfo{editor}{\bibfnamefont{J.~A.} \bibnamefont{Wheeler}} \bibnamefont{and}
  \bibinfo{editor}{\bibfnamefont{W.}~\bibnamefont{Zurek}}, eds.,
  \emph{\bibinfo{title}{Quantum Theory and Measurement}}
  (\bibinfo{publisher}{Princeton University Press},
  \bibinfo{address}{Princeton}, \bibinfo{year}{1983});
\bibinfo{author}{\bibfnamefont{W.~H.} \bibnamefont{Zurek}},
  \bibinfo{journal}{Rev. Mod. Phys.} \textbf{\bibinfo{volume}{75}},
  \bibinfo{pages}{712} (\bibinfo{year}{2003});
  \bibinfo{author}{\bibfnamefont{D.} \bibnamefont{Giulini}}
  \bibinfo{author}{\bibfnamefont{E.} \bibnamefont{Joos}}
  \bibinfo{author}{\bibfnamefont{C.} \bibnamefont{Kiefer}}
  \bibinfo{author}{\bibfnamefont{J.} \bibnamefont{Kupsch}}
  \bibinfo{author}{\bibfnamefont{I.~-0.} \bibnamefont{Stamatescu}}
  \bibinfo{author}{\bibfnamefont{H.~D.} \bibnamefont{Zeh}},
  \emph{\bibinfo{title}{Decoherence and the Appearance of a Classical World in
Quantum Theory}}
  (\bibinfo{publisher}{Springer-Verlag},
  \bibinfo{address}{Heidelberg}, \bibinfo{year}{1996});
\bibinfo{author}{\bibfnamefont{M.} \bibnamefont{Namiki}}
\bibinfo{author}{\bibfnamefont{S.} \bibnamefont{Pascazio}},
  \bibinfo{journal}{Phys. Rev. A} \textbf{\bibinfo{volume}{44}},
  \bibinfo{pages}{39} (\bibinfo{year}{1991}).


\bibitem[{\citenamefont{Tan and Walls}(1993)}]{TAW93}
\bibinfo{author}{\bibfnamefont{S.}~\bibnamefont{Tan}} \bibnamefont{and}
  \bibinfo{author}{\bibfnamefont{D.}~\bibnamefont{Walls}},
  \bibinfo{journal}{Phys. Rev. A} \textbf{\bibinfo{volume}{47}},
  \bibinfo{pages}{4663} (\bibinfo{year}{1993}).

\bibitem[{\citenamefont{Tegmark}(1993)}]{TEG93}
\bibinfo{author}{\bibfnamefont{M.}~\bibnamefont{Tegmark}},
  \bibinfo{journal}{Found. Phys. Lett.} \textbf{\bibinfo{volume}{6}},
  \bibinfo{pages}{571}
  (\bibinfo{year}{1993}).

\bibitem[{\citenamefont{Feynman et~al.}(1965)\citenamefont{Feynman, Leighton,
  and Sands}}]{FLS65}
\bibinfo{author}{\bibfnamefont{R.}~\bibnamefont{Feynman}},
  \bibinfo{author}{\bibfnamefont{R.}~\bibnamefont{Leighton}}, \bibnamefont{and}
  \bibinfo{author}{\bibfnamefont{M.}~\bibnamefont{Sands}},
  \emph{\bibinfo{title}{The Feynman Lectures on Physics}}, vol.
  \bibinfo{volume}{Vol. 3} (\bibinfo{publisher}{Addison-Wesley},
  \bibinfo{address}{Reading, MA}, \bibinfo{year}{1965}).

\bibitem[{\citenamefont{Schmiedmayer et~al.}(1995)\citenamefont{Schmiedmayer,
  Chapman, Ekstrom, Hammond, Wehinger, and Pritchard}}]{SCE95}
\bibinfo{author}{\bibfnamefont{J.}~\bibnamefont{Schmiedmayer}}
  \bibinfo{author}{~\bibnamefont{\textit{et. al.}}},
  \bibinfo{journal}{Phys. Rev. Lett.} \textbf{\bibinfo{volume}{74}},
  \bibinfo{pages}{1043} (\bibinfo{year}{1995});
\bibinfo{author}{\bibfnamefont{T.~D.} \bibnamefont{Roberts}},
  \bibinfo{author}{\bibfnamefont{A.~D.} \bibnamefont{Cronin}},
  \bibinfo{author}{\bibfnamefont{D.~A.} \bibnamefont{Kokorowski}},
  \bibnamefont{and} \bibinfo{author}{\bibfnamefont{D.~E.}
  \bibnamefont{Pritchard}}, \bibinfo{journal}{Phys. Rev. Lett.}
  \textbf{\bibinfo{volume}{89}} \bibinfo{pages}{200406 }(\bibinfo{year}{2002}).


\bibitem[{\citenamefont{Summhammer et~al.}(1987)\citenamefont{Summhammer, Rauch, and Tuppinger}}]{neutrons}
\bibinfo{author}{\bibfnamefont{J.} \bibnamefont{Summhammer}},
  \bibinfo{author}{\bibfnamefont{H.} \bibnamefont{Rauch}},  \bibnamefont{and}
  \bibinfo{author}{\bibfnamefont{D.} \bibnamefont{Tuppinger}},
  \bibinfo{journal}{Phys. Rev. A}
  \textbf{\bibinfo{volume}{36}} \bibinfo{pages}{4447
  }(\bibinfo{year}{1987});
\bibinfo{author}{\bibfnamefont{J.} \bibnamefont{Summhammer}}, \bibnamefont{and}
  \bibinfo{author}{\bibfnamefont{H.} \bibnamefont{Rauch}},
  \bibinfo{journal}{Phys. Rev. A}
  \textbf{\bibinfo{volume}{46}} \bibinfo{pages}{7284 }(\bibinfo{year}{1992}).




\bibitem[{\citenamefont{Chapman et~al.}(1995)\citenamefont{Chapman, Hammond,
  Lenef, Schmiedmayer, Rubenstein, Smith, and Pritchard}}]{CHL95}
\bibinfo{author}{\bibfnamefont{M.}~\bibnamefont{Chapman}}
  \bibinfo{author}{~\bibnamefont{\textit{et. al.}}},
  \bibinfo{journal}{Phys. Rev. Lett} \textbf{\bibinfo{volume}{75}},
  \bibinfo{pages}{3783} (\bibinfo{year}{1995}).

\bibitem[{\citenamefont{Berman}(1997)}]{BER97}
\bibinfo{editor}{\bibfnamefont{P.}~\bibnamefont{Berman}}, ed.,
  \emph{\bibinfo{title}{Atom Interferometry}}, Advances in Atomic, Molecular,
  and Optical Physics (\bibinfo{publisher}{Academic Press},
  \bibinfo{address}{San Diego}, \bibinfo{year}{1997}).

\bibitem[{\citenamefont{Kokorowski et~al.}(2001)\citenamefont{Kokorowski,
  Cronin, Roberts, and Pritchard}}]{KCR01}
\bibinfo{author}{\bibfnamefont{D.}~\bibnamefont{Kokorowski}}
  \bibinfo{author}{~\bibnamefont{\textit{et. al.}}},
  \bibinfo{journal}{Phys. Rev. Lett.} \textbf{\bibinfo{volume}{86}},
  \bibinfo{pages}{2191} (\bibinfo{year}{2001}).


\bibitem[{\citenamefont{Pritchard et~al.}(2001)\citenamefont{Pritchard, Cronin,
  Gupta, and Kokorowski}}]{PCK01}
\bibinfo{author}{\bibfnamefont{D.}~\bibnamefont{Pritchard}}
  \bibinfo{author}{~\bibnamefont{\textit{et. al.}}},
  \bibinfo{journal}{Ann. Phys.} \textbf{\bibinfo{volume}{10}},
  \bibinfo{pages}{35} (\bibinfo{year}{2001}).



\bibitem[{\citenamefont{Hornberger et~al.}(2003)\citenamefont{Hornberger,
  Uttenthaler, Brezger, Hackermuller, Arndt, and Zeilinger}}]{HUB03}
\bibinfo{author}{\bibfnamefont{K.}~\bibnamefont{Hornberger}}
  \bibinfo{author}{~\bibnamefont{\textit{et. al.}}},
  \bibinfo{pages}{160401}
  (\bibinfo{year}{2003}).

\bibitem[{\citenamefont{Brezger et~al.}(2002)\citenamefont{Brezger,
  Hackermüller, Uttenthaler, Petschinka, Arndt, and Zeilinger}}]{BJU02}
\bibinfo{author}{\bibfnamefont{B.}~\bibnamefont{Brezger}}
  \bibinfo{author}{~\bibnamefont{\textit{et. al.}}},
  \bibinfo{journal}{Phys. Rev. Lett.} \textbf{\bibinfo{volume}{88}}
  \bibinfo{pages}{100404}
  (\bibinfo{year}{2002}).


\bibitem[{\citenamefont{Sakurai}(1994)}]{SAK94}
\bibinfo{author}{\bibfnamefont{J.}~\bibnamefont{Sakurai}},
  \emph{\bibinfo{title}{Modern Quantum Mechanics}}
  (\bibinfo{publisher}{Addison-Wesley}, \bibinfo{address}{US},
  \bibinfo{year}{1994}).

\bibitem[{\citenamefont{Russek}(1960)}]{RUS60}
\bibinfo{author}{\bibfnamefont{A.}~\bibnamefont{Russek}},
  \bibinfo{journal}{Phys. Rev.} \textbf{\bibinfo{volume}{120}}
  \bibinfo{pages}{1536}
  (\bibinfo{year}{1960}).

\bibitem[{\citenamefont{Tellinghuisen}(1979)}]{TEL79}
\bibinfo{author}{\bibfnamefont{J.}~\bibnamefont{Tellinghuisen}},
  \bibinfo{journal}{J. Chem. Phys.} \textbf{\bibinfo{volume}{71}}
  \bibinfo{pages}{1283}  (\bibinfo{year}{1979});
\bibinfo{author}{\bibfnamefont{W.}~\bibnamefont{Baylis}}, \bibinfo{journal}{J.
  Chem. Phys.} \textbf{\bibinfo{volume}{51}}
  \bibinfo{pages}{2665}  (\bibinfo{year}{1969}).

\bibitem[{\citenamefont{Gustavson et~al.}(2000)\citenamefont{Gustavson,
  Landragin, and Kasevich}}]{GLK00}
\bibinfo{author}{\bibfnamefont{T.~L.} \bibnamefont{Gustavson}},
  \bibinfo{author}{\bibfnamefont{A.}~\bibnamefont{Landragin}},
  \bibnamefont{and} \bibinfo{author}{\bibfnamefont{M.~A.}
  \bibnamefont{Kasevich}}, \bibinfo{journal}{Classical and Quantum Gravity}
  \textbf{\bibinfo{volume}{17}}, \bibinfo{pages}{2385} (\bibinfo{year}{2000}).



\end{thebibliography}

\end{document}